# Nonlinear Effects on Quantum Interference in Electron Billiards

C. A. Marlow, R. P. Taylor, M. Fairbanks, and H. Linke

Physics Department, University of Oregon, Eugene OR 97403-1274, USA

**Summary**. Magnetoconductance fluctuations are used to study the effect of an applied bias on an electron billiard. At lower bias, nonlinear effects can be well described by electron heating alone, while at higher bias ($V > 2$mV, ~5% of the electron Fermi energy) non-equilibrium effects become significant. At high bias, we also observe that the spectral content of the MCF is sensitive to the nonequilibrium effects. Spectral behavior is consistent with a fractal scaling of the conductance fluctuations with magnetic field, resulting in the first observation of fractal conductance fluctuations outside of the linear regime of transport.

## 1 Introduction

In this work, we use electron quantum interference effects to study the effect of an applied bias on electron transport. The electron billiards used to study these effects were defined by e-beam lithography and wet etching of the two-dimensional electron gas (2DEG) formed in the GaInAs quantum well in the GaInAs/InP heterostructure (see Fig. 1(a)). A square (Figs. 1(b)) and rectangular (Fig. 1(c)) were studied with areas, after depletion, of 0.8 $\mu m^2$ and 3.4 $\mu m^2$, and Fermi energies of 35 meV and 38 meV, respectively. In both cases, the phase coherence length and mean free path were greater than the device dimensions resulting in phase-coherent, ballistic transport. Quantum interference effects lead to fluctuations in billiard conductance as a function of a perpendicular applied magnetic field, $B$. These magnetoconductance fluctuations, MCF, are a sensitive reproducible probe of the electron dynamics within the billiard[1] and will be used here to monitor the effect of an applied bias on electron transport.

In the presence of an applied bias, electrons are injected into the billiard with excess energy. If the electrons have time to thermalize before leaving the billiard, they relax through electron-electron scattering and the excess energy is distributed amongst the electrons in the billiard causing an in-crease in overall electron temperature inside the billiard. Previous experiments have found that for small applied bias voltages, in the µV range, the primary effect of the bias is electron heating.[2,3] The electron heating can be characterized by an effective temperature, $T_e(V)$, written:[2]

$$T_e(V) = \frac{T_L}{2} + \frac{1}{2}\sqrt{T_L^2 + e^{-\gamma \tau_\varepsilon(V)} \frac{3}{2\pi^2}\left(\frac{eV}{k}\right)^2} \qquad (1)$$

where $T_L$ is the temperature of the lattice, exp($-\gamma \tau_\varepsilon(V)$) is the fraction of electrons that thermalize before escaping the billiard, $\gamma$ is the escape rate,[4] and $\tau_\varepsilon(V)$ is the electron-electron interaction time. At the temperatures used here phase-breaking is dominated by electron-electron scattering, so the experimentally measured phase breaking length, $\tau_\phi$, will be used for $\tau_\varepsilon$ when calculating $T_e(V)$. $\tau_\phi$ was determined from the measured MCF using a well-established method that analyzes the correlation field of the fluctuations as a function of magnetic field.[5]

We use this heating model to study the importance of nonequillibrium effects in the mV range. MCF measurements were taken as a function of $T$ and $V$ and directly compared using Eq.1 to translate $V$ to $T_e(V)$; any departure between the two behaviors we interpret as nonequilibrium effects.

## 2 Experimental Results

The two-terminal magnetoconductance through the billiards was measured as a function of a perpendicular $B$ using a standard low frequency ac lock-in technique. In order to apply a bias across the billiard, a tuneable dc bias $V$ was added to a small ac signal (rms amplitude 20 µV on order of the thermal energy $kT \approx 20$ µeV). Measurements were made at a range of temperatures with $V = 0$ mV and also for a range of dc biases (up to 3 mV) at $T = 230$ mK.

Figure 1(d) shows the MCF for the square billiard measured for a range of $T$ (black curves) and $V$ (gray curves). The bias values have been related to the associated temperature using Eq. 1. At low bias, the fluctuations taken at a bias are similar to those at the corresponding temperature $T_e(V)$, consistent with previous observations in GaAs/AlGaAs billiards[3] where agreement was seen in the µV range. At higher bias, however, a departure is seen between the MCF measured at $V$ and those at the related $T$, indicating that at higher bias, the effect of the bias on the fluctuations is not just electron heating.

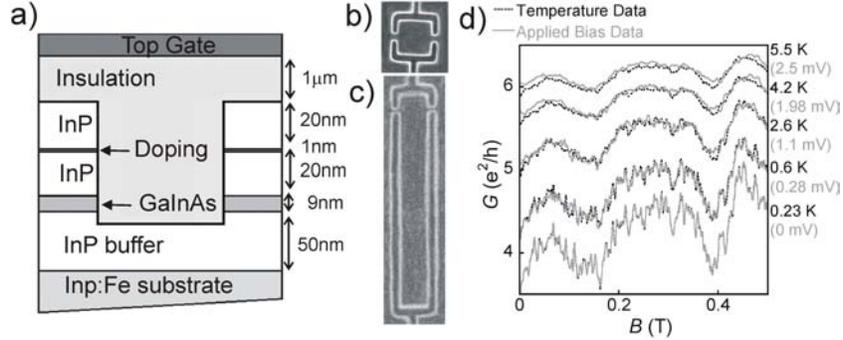

**Fig. 1.** a) Schematic representation of GaInAs/InP billiard system, scanning electron micrographs of the b) square and c) rectangular billiard and d) MCF for the square billiard, measured for a range of temperatures at zero bias (black curves) and range of bias at $T = 230$ mK (gray curves). Traces are offset for clarity.

We also investigate the effect of the bias on the spectral content of the fluctuations. The power spectrum, $S(f)$, of the MCF for the square billiard at $T = 0.6$ K and $V = 0$ mV is shown in Fig. 2(a). All MCF for the billiards presented here show $1/f^\alpha$ scaling, where $f = 1/\Delta B$ and $\alpha$ is the spectral exponent which characterizes the entire spectral content of the fluctuations. The $\alpha$ values observed indicate a fractal scaling[6] of the fluctuations consistent with previously observed fractal conductance fluctuations, FCF, in similar systems.[7]

Figures 2(b) and 2(c) show the dependence of the spectral exponent, $\alpha$, on both $T$ and $T_e(V)$ for the square and rectangular billiard respectively. We see that $\alpha$ increases with increasing $T$, consistent with previous results.[7] The same increase is seen with increasing $T_e(V)$. It appears, however, that at $T_e(V) \sim 4$ K, (corresponding to $\sim 2$ mV) the evolution of $\alpha$ with $V$ departs significantly from that of $\alpha(T)$, confirming again that the bias is having an additional effect besides heating.

## 3 Discussion

The deviations of the characteristics of the MCF indicate that the effect of the high bias is nontrivial; the bias does not just increase electron energy, but instead changes electron dynamics within the billiard. Future work needs to experimentally investigate the precise role of nonequillibrium electrons in the generation of MCF.

The fractal nature of MCF has been observed to be is robust to changes in many system parameters.[7] The effect of the nonequillibrium electrons on the spectral exponent is unexpected. Future exploration of this dependence may provide insight into the origins of FCF. In addition, further analysis of FCF in the nonlinear region may help uncover the mechanisms responsible for the symmetry breaking of the spectral content of the FCF with respect to reversal of magnetic field seen previously.[8]

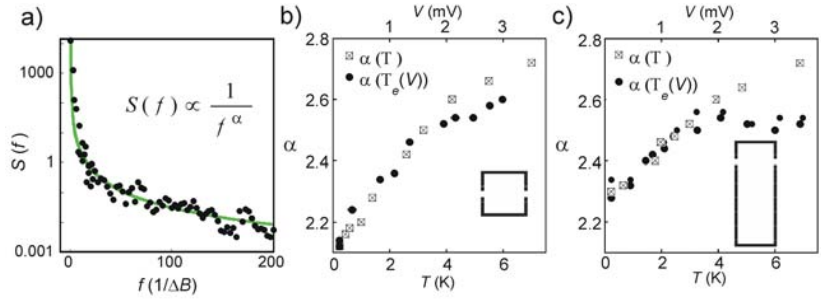

**Fig. 2.** a) Power spectra of MCF measured on square billiard ($T$ = 0.6 K, $V$ = 0 mV), spectral exponent $\alpha$ as a function of $T$ and $T_e$ ($V$) for b) square and c) rectangular billiard.

## References

1. Beenakker, C. W. J. and van Houten, H.: 'Quantum transport in semiconductor structures', *Solid State Physics*, edited by H Ehrenreich and D Turnbull, Academic Press, **44**, 1991.
2. Linke, H. et al.: 'Non-equilibrium electrons in ballistic quantum dot', Phys. Stat. Sol., **204**, 318, 1997.
3. Switkes, M. et al.: 'High bias transport and magnetometer design in open quantum dots',Appl. Phy. Lett., **72**, 471, 1998.
4. Jensen, R. V.: 'Chaotic scattering, unstable periodic orbits, and fluctuations in quantum transport',Chaos, **1**, 101, 1991.
5. Bird, J. P. et al.:'Phase breaking in ballistic quantum dots: a correlation field analysis', Surf. Sci., **361/362**, 730, 1996.
6. Barnsley, M. F. et al.: *The Science of Fractal Images*, Springer-Verlag, 1988.
7. Micolich, A. P. et al.:'Three key questions on fractal conductance fluctuations: Dynamics, quantization, and coherence', Phys. Rev. B **70**, 085302, 2004; Marlow, C. A. et al.: submitted, 2005.
8. Marlow, C. A. et al.: to be submitted, 2005.